\documentclass[default]{aastex631}

\accepted{May 6, 2023}

\begin{document}

\title{The hazardous km-sized NEOs of the next thousands of years}

\author[0000-0001-5875-1083]{Oscar Fuentes-Mu\~{n}oz}
\affiliation{Smead Department of Aerospace Engineering Sciences, University of Colorado Boulder, 429 UCB, Boulder, CO 80305-0429 USA}

\author[0000-0003-0558-3842]{Daniel J. Scheeres}
\affiliation{Smead Department of Aerospace Engineering Sciences, University of Colorado Boulder, 429 UCB, Boulder, CO 80305-0429 USA}

\author[0000-0003-0774-884X]{Davide Farnocchia}
\affiliation{Jet Propulsion Laboratory, California Institute of Technology, 4800 Oak Grove Dr, Pasadena, CA 91109, USA}

\author[0000-0001-9896-4585]{Ryan S. Park}
\affiliation{Jet Propulsion Laboratory, California Institute of Technology, 4800 Oak Grove Dr, Pasadena, CA 91109, USA}



\begin{abstract}
The catalog of km-sized near-Earth objects (NEOs) is nearly complete. Typical impact monitoring analyses search for possible impacts over the next 100 years and none of the km-sized objects represent an impact threat over that time interval. Assessing the impact risk over longer time scales is a challenge since orbital uncertainties grow. To overcome this limitation we analyze the evolution of the Minimum Orbit Intersection Distance (MOID), which bounds the closest possible encounters between the asteroid and the Earth. The evolution of the MOID highlights NEOs that are in the vicinity of the Earth for longer periods of time, and we propose a method to estimate the probability of a deep Earth encounter during these periods. This metric is used to rank the km-sized catalog in terms of their long-term impact hazard to identify targets of potential interest for additional observation and exploration.

\end{abstract}


\section{Introduction}

Asteroid impacts are one of the few natural disasters that can be prevented through human action. The main planetary defense efforts consist of observations, orbit determination and impact hazard assessment, and deflection/in-situ characterization. The near-Earth asteroid catalog is being completed by current and proposed surveys, providing new candidates of a future collision to study in more detail.

In 1998 the congress of the US requested NASA to detect and catalog 90\% of the km-sized NEO population\footnote{More details on the historical efforts of the U.S. Government to track and mitigate asteroids were given in two parts of a hearing before the Committee on Science, Space and Technology of Congress in March 19, 2013 and April 10, 2013. Full hearing statements accessible at {https://www.govinfo.gov/content/pkg/CHRG-113hhrg80552/pdf/CHRG-113hhrg80552.pdf}}. As of 2023-02-08, the catalog is around 95\% complete, with an estimated population of $962^{+52}_{-56}$ \citep{granvik_debiased_2018}. Impact monitoring systems estimate the orbits of newly discovered objects and compute any impact probabilities in future close encounters. Using the observational data available for a given object, the orbit is statistically estimated within an uncertainty region. This uncertainty region is efficiently sampled using various techniques to assess impact probabilities. 

The first generation impact monitoring system relied on the Line of Variations technique \citep{MILANI2005362}, sampling a suitably chosen direction of the uncertainty region. More recently, \cite{roa_novel_2021} describe a different approach that samples the full N-dimensional uncertainty region and identifies virtual impactors by using the impact condition as an observable. This latter approach is used by JPL's Sentry-II system\footnote{https://cneos.jpl.nasa.gov/sentry/}.

In this paper we investigate the potential impact risk over an order of magnitude larger timescales, in the next thousand years. To do this we review the two conditions required for an impact to occur \citep{Valsecchi2003}, and how the growth in orbit uncertainty affects them. The first one is that the Minimum Orbit Intersection Distance (MOID) has to be smaller than the combined radii of the two bodies, taking into account the gravitational focusing factor. This condition motivates the orbit condition for the definition of a Potentially Hazardous Asteroid (PHA): having an Earth MOID $<$ 0.05 au \citep{bowell1994earth}. Similarly, the MOID can be used to rule out NEOs for further potential impact analysis. The MOID is found as a function of the orbit elements of the Earth and those of the NEO, but does not directly depend on the position along the orbit \citep{Gronchi2005}. The uncertainty in these elements does not grow as fast as in mean anomaly, which allows us to propagate it confidently in longer timescales. Previous works studied the models required to propagate the MOID \citep{Gronchi2012}, including the applicability of the 3-body problem. In the presence of planetary encounters and complex long-term secular interactions, we must use numerical integration to propagate the orbits.

The second condition is on the timing of the flyby: the two bodies must be at the same time in the region in which their relative distance allows for a collision \citep{Valsecchi2003}. Uncertainty in the asteroids position grows faster in the direction of motion, limiting the assessment of future impacts. After a few centuries the uncertainty in mean anomaly can cover the whole orbit. This phenomenon is used as an assumption of analytical theories of impact rates in timescales of millions of years \citep{Opik1951,wetherill1967collisions}. In this work we keep track of the uncertainty in mean anomaly and use this assumption when the MOID condition is met. Previous works combine these assumptions in hundreds of thousands of years timescales \citep{Vokrouhlicky2012,Pokorny2013}, using analytical models of the long-term dynamics. In these much longer timescales the uncertainty in NEO orbits grows large enough that lower fidelity models of the long-term dynamics provide good estimates of the frequency of close encounters \citep{Fuentes-Munoz2021}. Thus, in the shorter timescales of this work, we propose the combination of the two conditions although we propagate the orbits of the NEOs numerically.

We investigate the long-term MOID dynamics and identify km-sized NEOs that are frequently in the neighborhood of Earth. Then, we keep track of the uncertainty in mean anomalies during those low-MOID periods of the NEOs. Using the analytical approximations we estimate the probability of a deep encounter. This metric allows us to rank the km-sized NEO population, highlighting a few km-sized NEOs for further detailed analysis.

\newpage
\section{Long-term NEO Hazard characterization}

In this section we describe the tools and methods used to analyze the long-term dynamics of the km-sized NEO population and the estimation of their potential impact hazard. The MOID time histories are obtained following the propagation of the orbit. Hence, we first describe the orbit propagator as motivated by the NEO long-term dynamics and then the MOID algorithm and dynamics. Last, we introduce the long-term collision hazard metric that is used to rank the selected group of near-Earth objects.

\subsection{Orbit propagation}

The orbits of the NEOs are propagated using the JPL small-body integrator which is based on an N-body model that includes Sun, planets, Pluto, Moon and small-body perturbers \citep{Farnocchia2015-IV}. When the Yarkovsky effect was detected from astrometric data \citep{Farnocchia2013-2}, we added it to the force model. The ephemeris models used in the integration are DE441 \citep{Park2021TheDE441} for the planets and SB441-N16 for the largest main-belt bodies\footnote{Available at: \url{ftp://ssd.jpl.nasa.gov/pub//eph/small_bodies/asteroids_de441/SB441_IOM392R-21-005_perturbers.pdf}}.

\begin{figure}[htb!]
	\centering
	\includegraphics[scale=1]{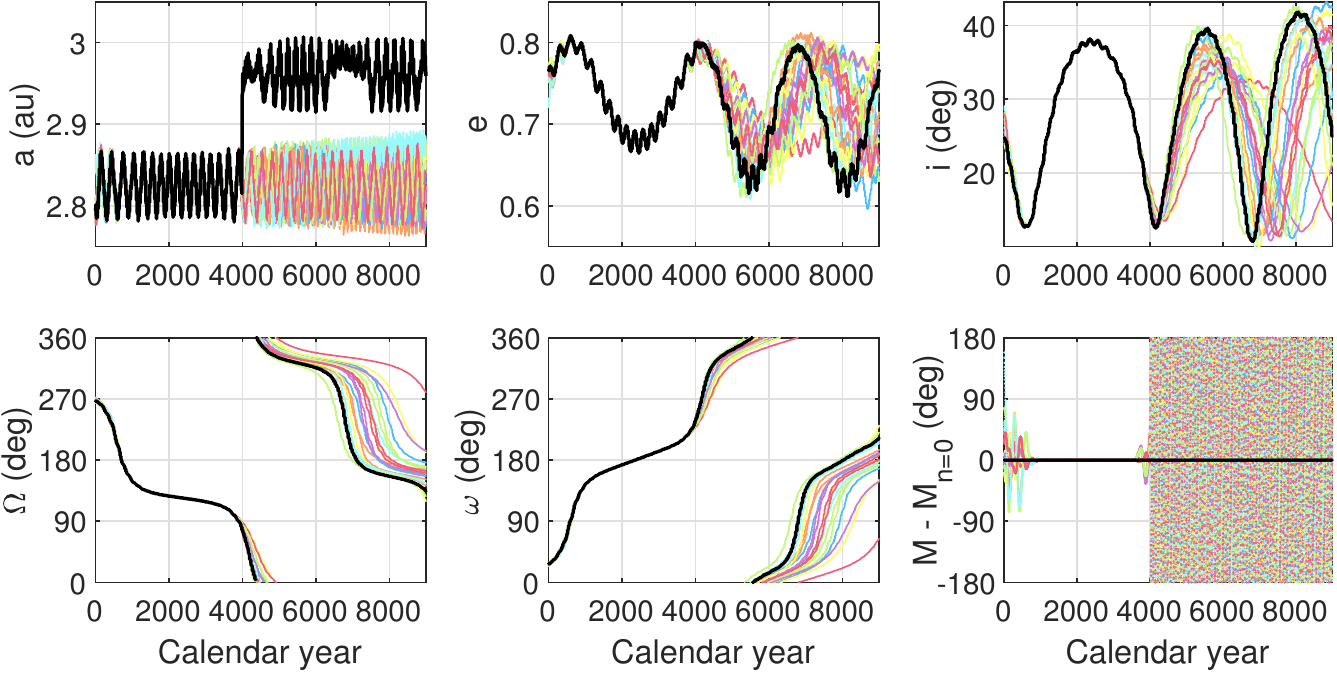}
	\vspace{10mm}
	\includegraphics[scale=1]{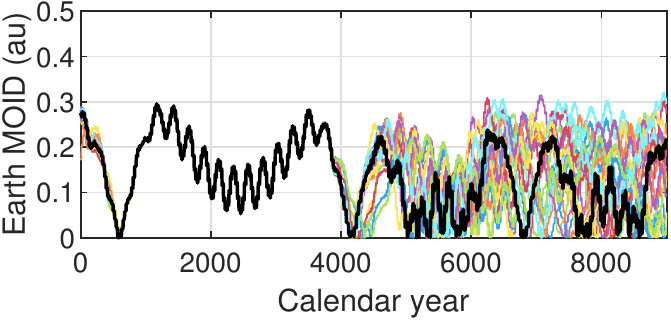}
	\hspace{3mm}
	\includegraphics[scale=1]{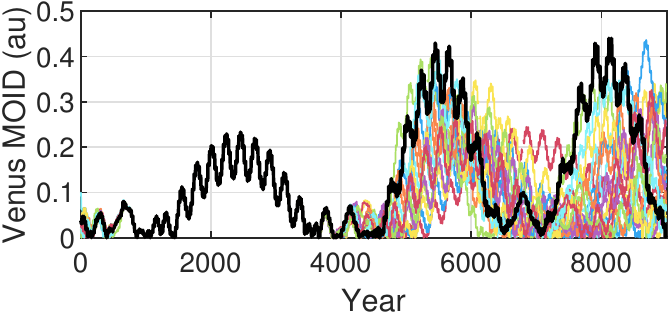}
	
	\caption{Numerical propagation of the orbit of 2015 FP332, a km-sized NEO. The trajectories are shown using the Keplerian elements semi-major axis, eccentricity, inclination, longitude of the ascending node, argument of perihelion and mean anomaly with respect to the nominal propagation. Individual Monte-Carlo runs (N=21) are shown in colors, the nominal trajectory is shown in a black line. The bottom rows show the propagation of Earth and Venus MOID.}
	\label{fig:prop-examp}
\end{figure}

Figure \ref{fig:prop-examp} shows the propagation of the orbit of 2015 FP332, which reveals the relevant dynamical effects to the long-term dynamics of near-Earth objects. 2015 FP332 is in a Lidov-Kozai cycle \citep{LIDOV1962719,kozai1962secular}, in which periods of high eccentricity are exchanged with periods of high inclination. In this case, both longitude of the node and argument of perihelion drift secularly.

Planetary encounters can cause the exponential growth of the distance between initially neighboring trajectories, a necessary condition for chaos \citep{tancredi_chaotic_1998}. Neighboring trajectories of near-Earth asteroids can diverge in timescales ranging from decades, such as 99942 Apophis \citep{farnocchia_yarkovsky-driven_2013}; to hundreds of years, such as 29075 (1950 DA) \citep{FARNOCCHIA2014321}; to tens of thousands of years, such as 433 Eros \citep{michel1996orbital}. In this process the linear approximation of the state uncertainty can quickly become inaccurate. 

{Figure \ref{fig:prop-examp} shows the propagation of the multiple samples or virtual asteroids of the orbit of 2015 FP332. In this example,} the nominal trajectory of 2015 FP332 experiences a very close Venus encounter that causes the rapid increase in semi-major axis. Once each initially neighboring virtual asteroid diverges to a different trajectory it experiences a unique sequence of close encounters. This effect motivates the use the MOID to estimate long-term probabilities of collision. The resulting dynamics under these encounters are very nonlinear, and the orbits of near-Earth objects in these timescales become stochastic. For this reason, we sample the uncertainty in the orbits of NEOs and propagate them in a Monte Carlo simulation. The detection of potential impactors of {small} probabilities is out of the scope of this work, in which the main metric of interest is the MOID. For this reason we run a limited number of Monte Carlo samples (N=21), which allows us to distinguish the main dynamical effects as well as the uncertainty in mean anomaly.

The presence of close encounters is expected if the near-Earth object has a small MOID with any of the planets. Thus, tracking the evolution of the MOID is not only relevant for the evaluation of the probability of collision with Earth but to understand when the dynamics are subject to nonlinear stochastic variations. The evolution of the orbit of 2015 FP332 in Figure \ref{fig:prop-examp} shows the effect of a low-MOID period in the long-term prediction. A Venus low-MOID enables close approaches that cause the rapid expansion of the distribution of orbits and mean anomaly to become unknown.

\subsection{MOID algorithm and dynamics}

The MOID is the result of the optimization of the relative distance between two bodies over their respective fast angles. There are multiple algorithms to compute the MOID in the literature, including analytical methods \citep{Gronchi2005} and numerical methods such as \cite{Wisniowski2013} or \cite{HedoPelaez}, which is used in this work. The MOID, a function of the osculating orbit elements, is then computed when post-processing the numerically integrated trajectory.

\begin{figure}[htb!]
	\centering
	\includegraphics[width=2.8in]{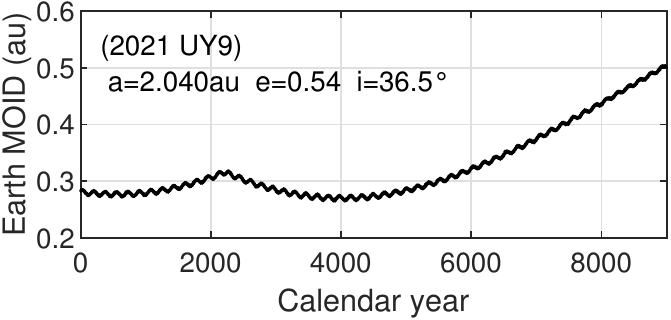}
	\hspace{3mm}
	\includegraphics[width=2.8in]{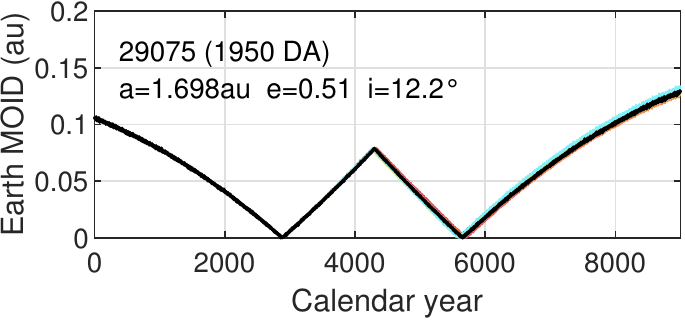}
	\vspace{3mm}
	\includegraphics[width=2.8in]{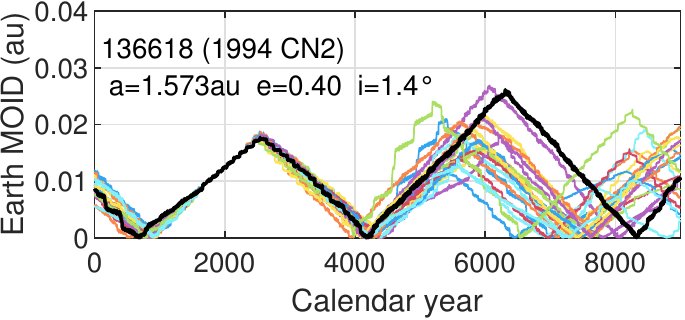}
	\hspace{3mm}
	\includegraphics[width=2.8in]{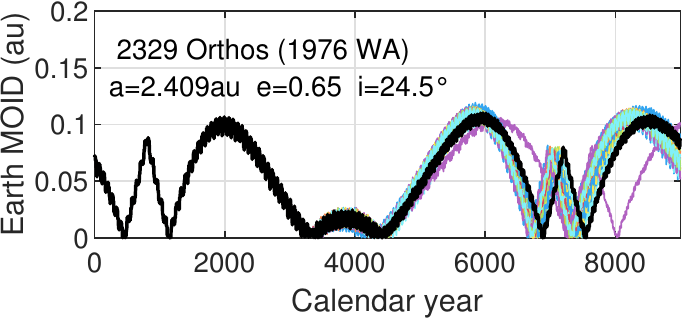}

	\caption{Propagation of the Earth MOID of a few selected examples of km-sized near-Earth objects. For each asteroid we show the 21 Monte Carlo simulations (colors) and the nominal orbit propagation (black, continuous). The orbit elements in the figure indicate the initial conditions of the propagation.}
	\label{fig:moid-examples}
\end{figure}

Depending on the dynamical effects on the asteroid described in the previous section we find a variety of MOID long-term dynamics trajectories. Figure \ref{fig:moid-examples} shows a few examples of the dynamics of the propagation of the MOID for four km-sized NEOs. 2021 UY9 represents the simplest case, in which the MOID does not become small throughout the simulation time therefore making Earth impacts impossible. The uncertainty in the orbit of 2021 UY9 remains small throughout the propagation. The case of 29075 (1950 DA) is the very common case for NEOs, in which the MOID drifts secularly until a future zero crossing that lasts a short period of time, of about a century. The example of 136618 (1994 CN2) is similar to the case of 2015 FP332 in Figure \ref{fig:prop-examp}, in which the date in which the MOID becomes small for the first time is uncertain. After a low-MOID period the trajectory becomes more uncertain. The last example, 2329 Orthos (1976 WA), illustrates the scenario of an extended period of time with a low Earth MOID. This is caused by the combination of two effects, a large amplitude of short-period oscillations and a favorable phasing of the secular cycle. 

{The examples of Figure \ref{fig:moid-examples} are also representative of the growth in uncertainty of the MOID. In the two top cases the uncertainty remains small for thousands of years. The km-sized NEA population have well defined orbits and their Earth MOIDs remain well known for hundreds of years. Previous works focus on mapping the orbit covariance into a confidence region of the MOID \citep{Gronchi2007}. However, the uncertainty in the orbit can become far from Gaussian in long-term orbit propagations. Thus, we use statistics of the Monte Carlo propagation as indicators of the spread of the distribution as well as the confidence in our predictions.}

\subsection{Long-term impact probability estimation}\label{s:metric}

The complexity in long-term MOID dynamics that we showed in the previous section motivates the development of a systematic method to quantify the long-term Earth impact hazard of NEOs. We propose a novel metric to characterize the potential impact hazard that consists in an estimated probability of collision {between planet and NEO}. The probability of collision is a problem primarily studied in two major timescales. The fundamental problem of impact hazard assumes the position of the asteroid within its orbit is reasonably well determined and it is possible to precisely determine the geometry of the subsequent close encounters. In the case of potentially hazardous asteroids, this analysis can typically be completed for one or two centuries  \citep{chamberlin_sentry_2001,roa_novel_2021}. In these timescales the uncertainty in the orbit of many NEOs starts to grow large enough that the position within the orbit can become unknown. This effect motivates the statistical assumption of a uniformly distributed mean anomaly.

Traditional impact probability theories assume that the orbit elements of the two objects involved are constant and have one intersection point \citep{Opik1951,wetherill1967collisions}. Then, the probability depends on the timing of the orbits, which is when the mean anomalies are assumed uniformly distributed. This timing probability, here $P_{MA}$, is the probability that both bodies are in the right time at the right place, i.e., the range of mean anomalies that corresponds to a collision or flyby within a small distance.

There are a few options for the timing probability in the literature. In the most simplified case, we can assume the planet's orbit to be circular with \"{O}pik's formula  \citep{Opik1951}, in which the probability is function of the Keplerian elements of the asteroid a-e-i.  \cite{wetherill1967collisions} then derived an expression for an elliptic orbit of the planet, with the problem of being singular at zero inclination. In this work we use this expression as re-derived recently in  \cite{JeongAhn2017} for regular non-tangential encounters. 
In particular, we use the extended expression for the case in which the two objects do not exactly intersect. That means that the MOID is a positive value between 0 and the distance threshold for the close encounters of interest $d$. Thus, the probability that two objects have a close encounter with closest approach distance smaller than $d$ is:

\begin{equation}\label{eq:PMA}
    P_{MA} = \frac{2Ud}{T_p T_{NEO}|\bf{v}_{p} \times \bf{v}_{NEO}|} \sqrt{1 - \frac{MOID^2}{d^2}}
\end{equation}

\noindent
where $\bf{v}_p$ and $\bf{v}_{NEO}$ are the velocities of the planet and the asteroid at the point that defines the MOID, $U$ is the relative velocity at the same point, and $T_p,T_{NEO}$ are the respective orbit periods. {The square root term of equation \ref{eq:PMA} adjusts the probability for a non-zero MOID. If MOID $>d$, the probability is assumed to be zero.} This expression can be averaged for a MOID uniformly distributed between 0 and $d$. However, in this work we do not need to make this assumption as we keep track of the MOID throughout our long-term propagation and the distribution can be far from uniform in the range $0< $MOID$ \leq d$. 

Once we allow the orbit of the NEO to be time-varying, we can obtain the probability of collision as the combination of two terms: the probability that there is an intersection between the planet and near-Earth object and $P_{MA}$. If we investigate a potential {Earth} collision, the condition is that the Earth MOID is smaller than the combined radii of the two bodies considering gravitational focusing as required. The gravitational focusing factor virtually extends the radius of the planet to account for trajectories that lead to a collision due to the planet's gravity, and is a function of the incoming velocity of the asteroid $V_\infty$ and the mass and radius of the planet $M_p,R_p$:

\begin{equation}\label{eq:grav focus}
    \gamma = \sqrt{1 + \frac{2GM_p}{R_p V_\infty^2}}
\end{equation}

This approach has been used in the past to obtain the probability of collision for asteroids under the Lidov-Kozai cycle \citep{Vokrouhlicky2012,Pokorny2013}. In that case, the generalized probability of collision is obtained as the sum over all the crossing configurations (noted with *) of the fraction of time that the NEO spends withing the distance threshold times the timing probability:

\begin{equation}
    P = \sum_{*} \left( \frac{\Delta t_{MOID<d}}{T_{sec}}\right)^{*}  P_{MA} (d,  K_{p}^{*} , {K}_{NEO}^{*}) 
\end{equation}

\noindent
where $K_{P},K_{NEO}$ are the Keplerian elements of the planet and the NEO and $\Delta t_{MOID<d}$ is the amount of time that the NEO spends within the distance threshold $d$. \cite{Vokrouhlicky2012} and \citep{Pokorny2013} model the long-term asteroid dynamics with an analytical solution of the Lidov-Kozai cycle of the Jupiter perturbation, which defines the secular period $T_{sec}$ as the Lidov-Kozai period. As a result, the fraction $\Delta t$ and intersection configurations are computed analytically. As we show in the previous section, defining the times in which the MOID is small can be a complex problem under a wide range of dynamical contributions. In this work we propagate the orbit numerically to find the low-MOID periods. Because there is not a small discrete number of crossings along the long-term dynamics of the NEO, we estimate the probability as the average throughout the propagation time $T$ using equation \ref{eq:Pkappa}.

\begin{equation}\label{eq:Pkappa}
    P = \frac{1}{T} \int_{T} \kappa  P_{MA} (d,  K_{{p}} , {K}_{NEO}) dt
\end{equation}

This integral is computed numerically using the numerically integrated trajectories. The factor $\kappa$ is introduced so we can null the contribution of the trajectory in which the position of the object is deterministic within its orbit, i.e., when the uncertainty in mean anomaly is small. {$\kappa$ is set to 0 before the first date in which we find that the standard deviation in mean anomalies is larger than 10 degrees, and set to 1 elsewhere.} This distinction allows us to rule out the associated risk of objects that currently have a very low MOID but their position is properly constrained for the duration of their visit to {the planet}'s vicinity. {In addition, we check the close encounters that were recorded in the propagation before we can use the analytical expression for $P_{MA}$ in equation \ref{eq:PMA}.}

\newpage
\section{Km-sized NEO population long-term characterization}

We analyze the potential impact hazard of the km-sized NEO population in the next millennium. Using the very low-MOID necessary condition for a potential collision, we can rule out the collision hazard when this condition is not met. Then, considering the statistical evolution of the mean anomalies, we rank this group of NEOs depending on their long-term implicit impact hazard.

\subsection{NEOs frequently in Low-MOID regions}\label{sec:MOIDprop}

The orbits of the known km-sized NEO population are propagated starting from their orbit solutions in JPL's Small-body Database as of 2022-09-15\footnote{Small-body Database available for query at: \url{ssd.jpl.nasa.gov/tools/sbdb_query.html} }. For each NEO we find the first date at which a low-MOID period is found between all of the Monte Carlo samples, with a threshold defined as MOID $<0.01$ au (235 Earth Radii or 3.89 Lunar Distances). At this threshold, the incoming velocity $V_\infty$ required for a collision is of $0.05$ km s$^{-1}$ or less, from solving equation \ref{eq:grav focus}. From a statistical point of view, this relative velocity is extremely unlikely\citep{Farnocchia_2021,HARRIS2021114452}. 



\begin{figure}[htb!]
	\centering
        \includegraphics[scale=1]{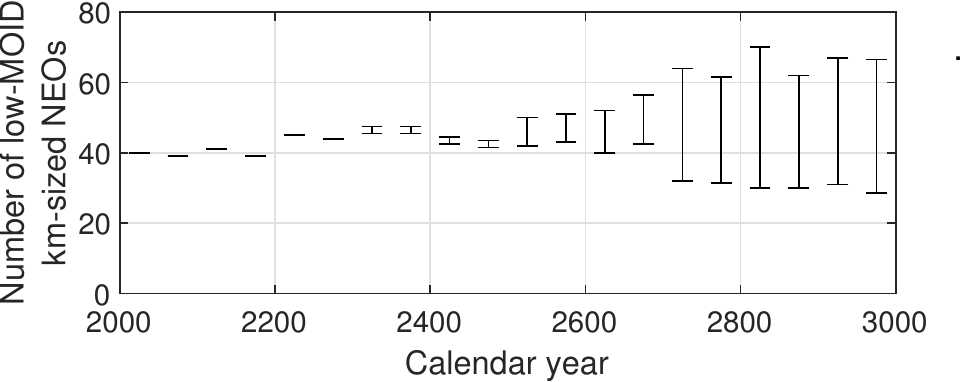}
        	
    \caption{Number of km-sized NEOs that present a low-MOID throughout the next 1000 years. defined as an Earth MOID $<$0.01au. Minimum number of NEOs is estimated by NEOs in which there was an agreement between all Monte Carlo runs. Maximum number of NEOs is estimated by at least one Monte Carlo run fulfilling the low-MOID condition.}
	\label{fig:hists_t0}
\end{figure}

\begin{figure}[htb!]
	\centering
	\includegraphics[scale=1]{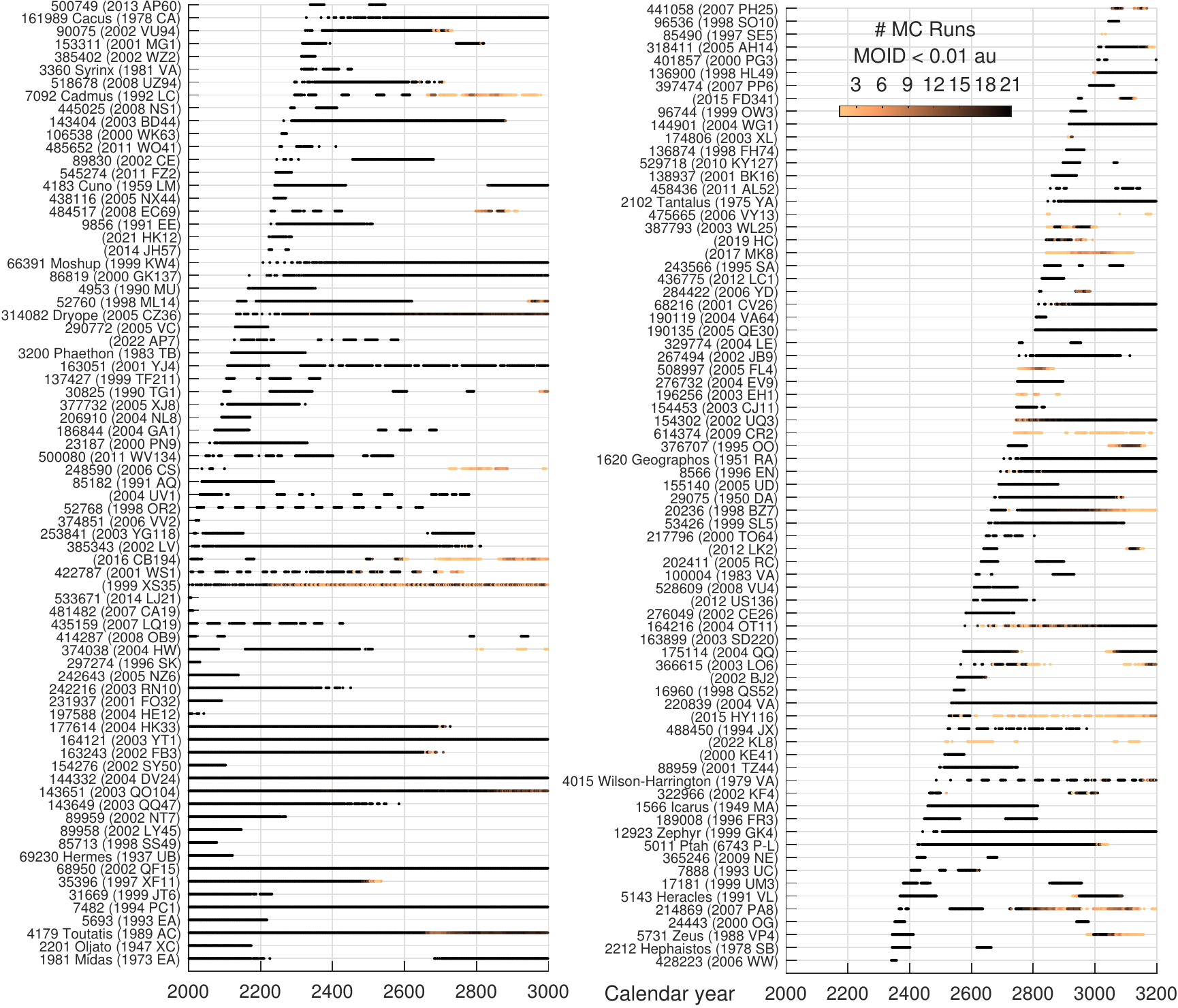}
        
	\caption{Km-sized NEOs that meet the MOID $<0.01$ au condition in the next 1000 years. Color code indicates the number of Monte-Carlo samples that show a MOID $<$ 0.01 au at the given date. Black means an agreement between all Monte Carlo runs to show a low-MOID. The NEOs are sorted by the first date in which they meet the MOID $<$ 0.01 au condition.}
	\label{fig:km-pop}
\end{figure}

The first question we answer is how many km-sized NEOs currently have a MOID $<0.01$ au, and how will this number evolve in the next 1000 years. As of the time in which the JPL's SBDB was queried, there are 40 NEOs that fulfill this condition. The evolution of this estimated number of NEOs is shown in Figure \ref{fig:hists_t0}. As the uncertainty in the orbits of the NEOs grows into the future, only some of the MC samples may have MOID $<0.01$ au. This phenomenon is shown in more detail in Figure \ref{fig:km-pop}, which shows the estimated range of km-sized NEOs with MOID $<0.01$ au based in the Monte Carlo samples. The uncertainty remains very small ($\pm$1 body) throughout the next 500 years. By the end of the millennium, this number is in the range of 26-72 km-sized NEOs. As mentioned earlier, none of these objects pose a collision threat to Earth in the next 100 years.


Individual results of the MOID propagation are shown in Figure \ref{fig:km-pop}. We sorted the km-sized NEOs by the date in which they meet the MOID $<0.01$ au condition. As defined by the length of their low-MOID periods, we show NEOs that are expected to be continuously in the vicinity of Earth as opposite to the ones that are for a brief period of time. We observe that even if the number of NEOs will never exceed an average value of ~40-45 NEOs, the total number of unique low-MOID NEOs in the next 1000 years is of almost 150.

\newpage
We list the km-sized NEOs that frequently experience MOID $<0.01$ au in Table \ref{tab:top10-frac} based in the fraction of the next 1000 years that they meet the condition. Their implicit probability of collision is assessed in the next section. There are 4 objects whose MOID remains lower than 0.01 au throughout this millennium: 7482 (1994 PC1), 68950 (2002 QF15), 164121 (2003 YT1), 144332 (2004 DV24). In the second and third case, the mean anomaly remains well defined throughout the millennium.

\begin{figure}[htb!]
 \centering
 	\includegraphics[width=2.8in]{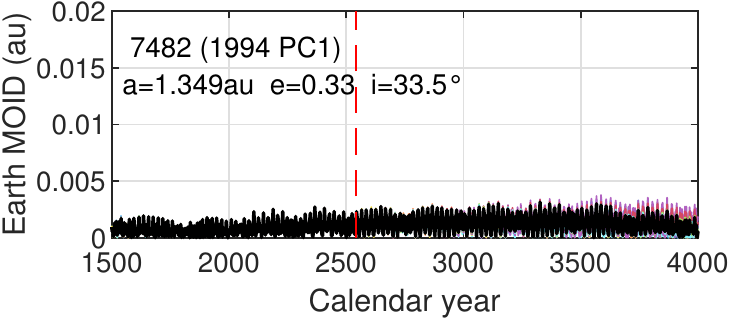}
	\hspace{3mm}
	\includegraphics[width=2.8in]{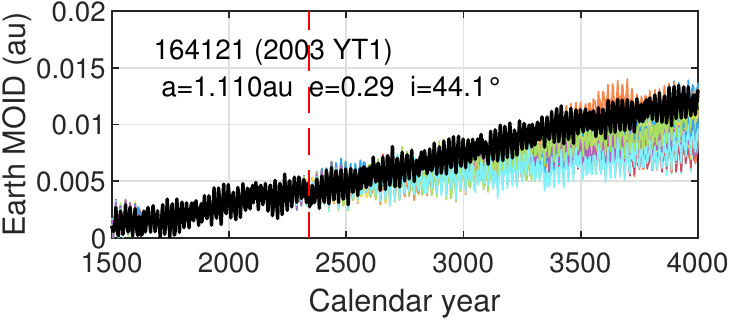}
	\vspace{3mm}
	\includegraphics[width=2.8in]{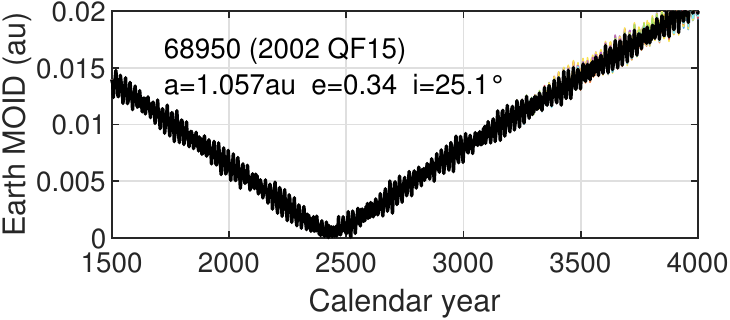}
	\hspace{3mm}
	\includegraphics[width=2.8in]{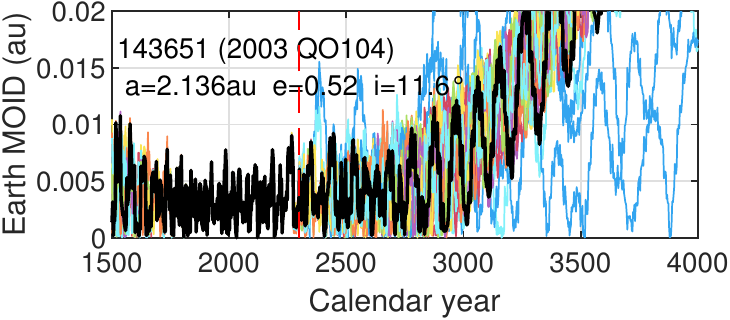}

	\caption{Propagation of the Earth MOID of km-sized NEOs with a low-MOID for a large fraction of the next 1000 years, as shown in Table \ref{tab:top10-frac}. Individual Monte-Carlo runs are shown in colors, black continuous line shows the nominal trajectory. The dashed red line indicates the first date in which the standard deviation in mean anomaly was found greater than 10 degrees, which does not happen for 68950 (2002 QF15).}
	\label{fig:moid-frecs}
\end{figure}

The propagation of the MOID of a few of the top-ranked NEOs is shown in Figure \ref{fig:moid-frecs}. Most remarkably, we can see how low the MOID of 7482 (1994 PC1) persists throughout the next 1000 years. In 68950 (2002 QF15) and 164121 (2003 YT1) we observe a secular drift in the MOID. In the case of 68950 (2002 QF15), this secular drift predicts a near-zero MOID around year 2500. In the case of 164121 (2003 YT1), the MOID is increasing at a relatively slow rate. The last example, 143651 (2003 QO104), shows a large amplitude of the MOID around zero, which motivates additional analysis of the long-term hazard.

\begin{table}[htb]\label{tab:top10-frac}
\caption{10 km-sized NEOs with the largest fraction of time with low MOID over the next 1000. Time fraction indicates the average fraction of time with MOID $<$ 0.01 au among the Monte Carlo experiments. The first date of standard deviation in Mean Anomaly $>10^{\circ}$ is shown with initial orbit elements at the ephemeris retrieval date, 2022-09-15. $V_\infty$ is the relative velocity at the first time that MOID $<$ 1 LD of the nominal solution.}
\begin{tabular}{ccccccc}
\hline
NEO                 & $\Delta t/T$ & $T_{S>10^{\circ}}$ & a (au) & e      & i (deg) & $V_\infty$ (km s$^{-1}$) \\ \hline
  7482 (1994 PC1) & 1.000 & 2541 & 1.349 & 0.330 & 33.47 & 19.68\\ 
 68950 (2002 QF15) & 1.000 & 3288 & 1.057 & 0.344 & 25.15 & 16.06\\ 
144332 (2004 DV24) & 1.000 & 3285 & 1.423 & 0.290 & 55.90 & 29.83\\ 
164121 (2003 YT1) & 1.000 & 2341 & 1.110 & 0.292 & 44.06 & 23.71\\ 
143651 (2003 QO104) & 0.945 & 2297 & 2.136 & 0.524 & 11.61 & 9.72\\ 
  4179 Toutatis (1989 AC) & 0.927 & 2516 & 2.545 & 0.625 & 0.45 & 12.19\\ 
314082 Dryope (2005 CZ36) & 0.750 & 2352 & 2.238 & 0.575 & 16.14 & 14.05\\ 
 86819 (2000 GK137) & 0.744 & 2565 & 1.996 & 0.506 & 10.06 & 10.07\\ 
385343 (2002 LV) & 0.740 & 2960 & 2.315 & 0.605 & 29.53 & 20.14\\ 
177614 (2004 HK33) & 0.702 & 3507 & 1.888 & 0.521 & 5.44 & 11.37\\  \hline
\end{tabular}
\end{table}

\newpage
\subsection{Upcoming hazardous km-sized NEOs}

In the previous section we inspected the necessary condition for very close encounters to occur: a MOID $<0.01$ au. The next step is to estimate the collision probability by making assumptions on their mean anomalies. The method to compute this timing probability once the MOID is low was described in section \ref{s:metric}. To study the potential impact hazard we set a smaller close approach threshold ($1$ LD) and take into account the deterministic parts of the NEO position during the orbit propagation. In addition, we study the list of close approaches generated in the Monte Carlo experiment to validate our predictions.

The analytical expressions for the probability of collision assume uniformly distributed mean anomalies of the bodies. The initial conditions of the propagation start from a well defined mean anomaly of the NEOs. Thus, we need to track the evolution of the uncertainty in mean anomaly to know when we can start using the analytical estimates. Using our Monte Carlo experiments we compute the standard deviation in mean anomaly separation from the nominal trajectory.

\begin{figure}[h!]
	\centering
	\includegraphics[scale=1]{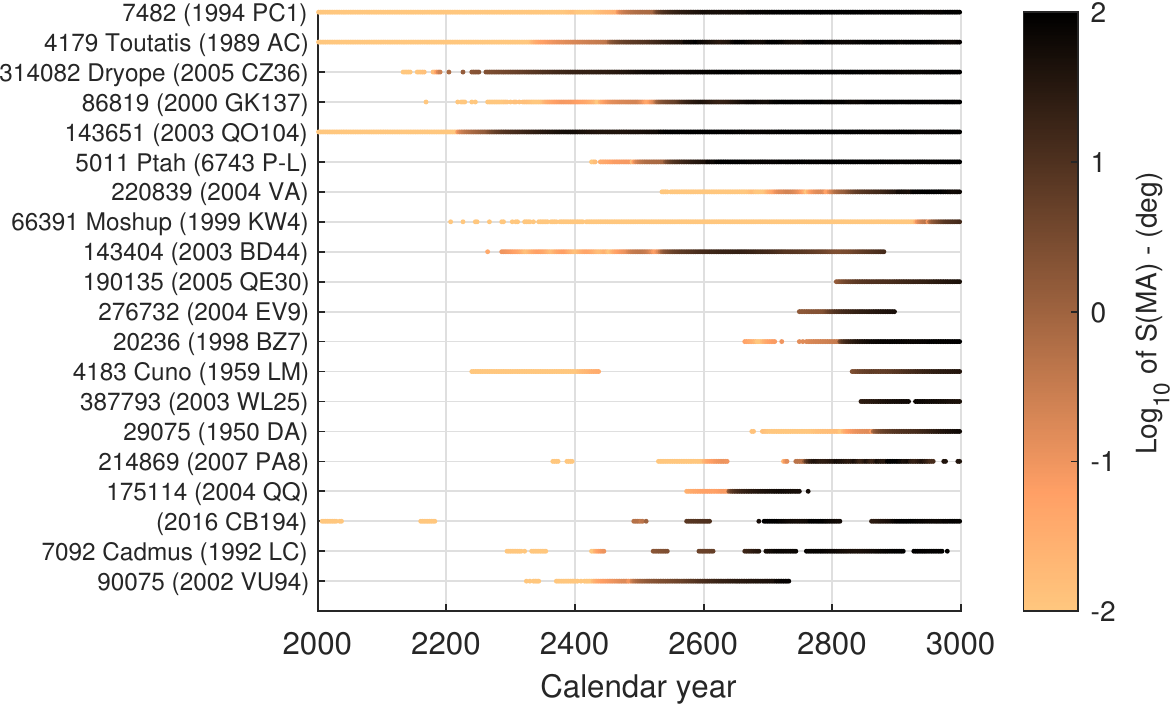}

	\caption{Km-sized NEOs with non-zero estimated probability of encounters closer than 1 LD. Color code indicates the standard deviation in mean anomaly, S(MA), in log10. S(MA) is shown only in dates in which the MOID is lower than 0.01 au.}
	\label{fig:upcoming-MA}
\end{figure}

\newpage
We propagate the orbits of the km-sized NEO population for 1000 years and study when the MOID is smaller than a Lunar Distance (1 LD). When the standard deviation in mean anomaly is large, we estimate the probability of close encounters. In Figure \ref{fig:upcoming-MA} we list the km-sized NEOs showing the dates in which we found a low-MOID and sorted by their estimated probability of close encounters. The low-MOID regions are color coded with the standard deviation in mean anomaly. The combination of this information highlights the future periods of time in which the position of the NEOs is unknown.

\begin{figure}[htb!]
	\centering
	\includegraphics[width=2.8in]{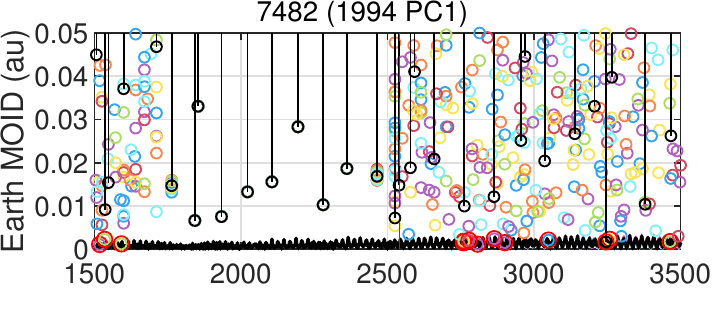}
	\hspace{3mm}
	\includegraphics[width=2.8in]{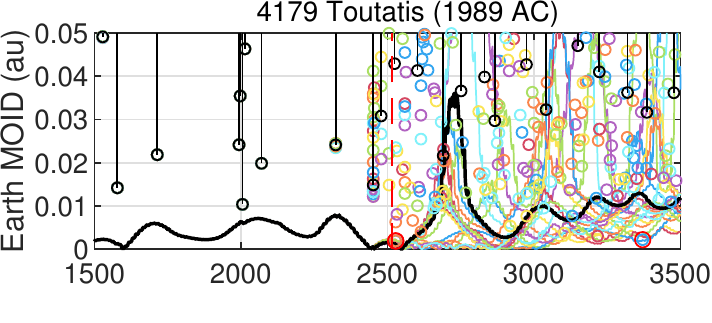}
	\includegraphics[width=2.8in]{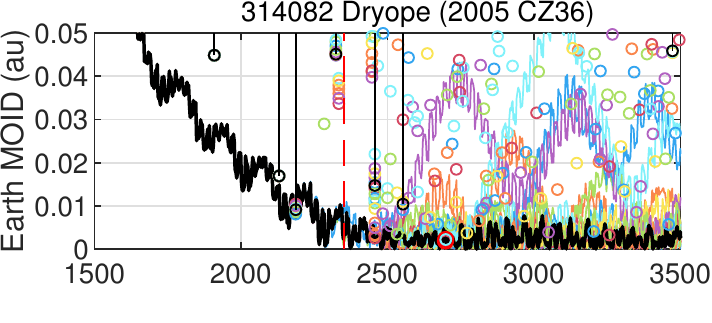}
	\hspace{3mm}
	\includegraphics[width=2.8in]{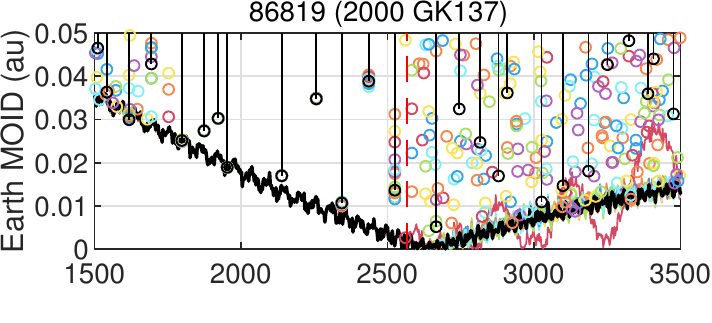}
	\includegraphics[width=2.8in]{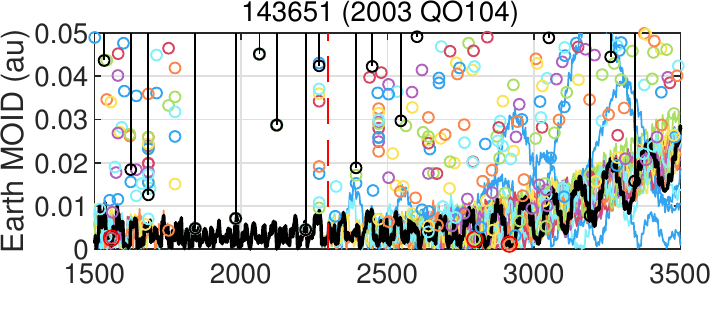}
	\hspace{3mm}
	\includegraphics[width=2.8in]{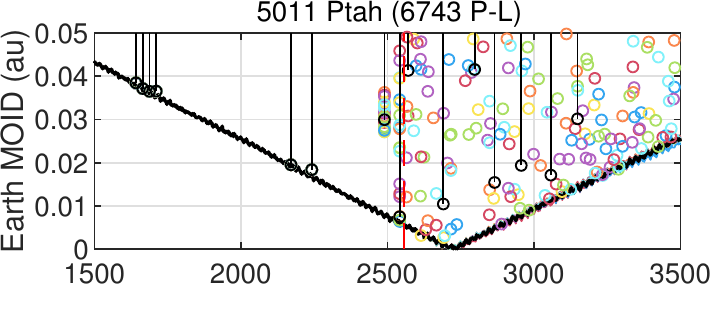}
	\includegraphics[width=2.8in]{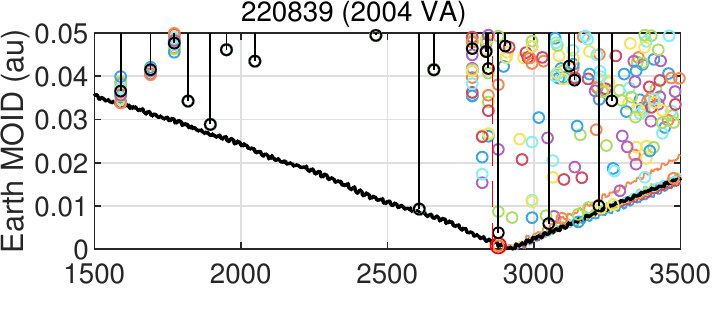}
	\hspace{3mm}
	\includegraphics[width=2.8in]{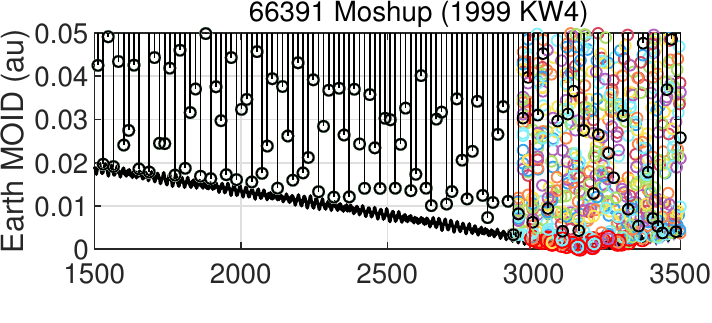}
	\includegraphics[width=2.8in]{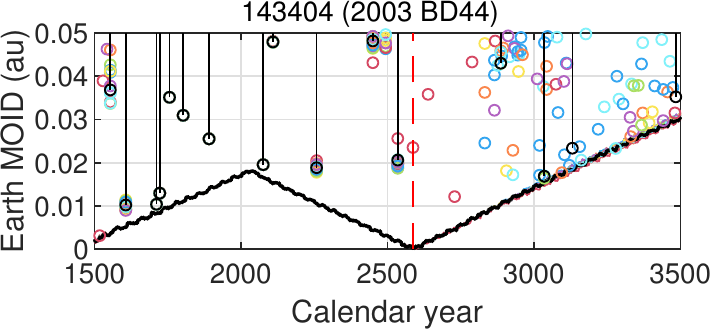}
	\hspace{3mm}
	\includegraphics[width=2.8in]{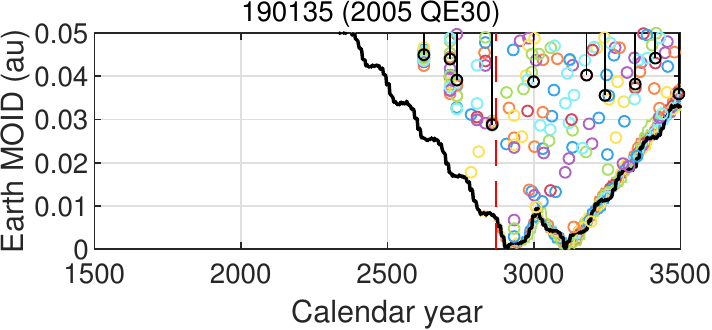}

	\caption{Propagation of the Earth MOID of km-sized NEOs with a non-zero probability of having an encounter closer than 1 LD by year 3000. Individual Monte-Carlo runs are shown in colors, black continuous line shows the nominal trajectory. The dashed red line indicates the first date in which the standard deviation in mean anomaly was found greater than 10 degrees. Close encounters are indicated with circles in colors and close encounters of the nominal trajectory are shown as vertical black nodes. Encounters of closest approach distance $< 1 LD$ (0.0026 au) are highlighted with a larger red circle.}
	\label{fig:moid-hazardous-non0s}
\end{figure}

\begin{table}[htb!]\label{tab:top10-hazLD}
\caption{ 28 km-sized NEOs with non-zero estimated probability of a deep encounter ($d_{CA}<$1 LD) in the next 1000 years, {as averaged over Monte Carlo runs}. Time fraction indicates the average fraction of time with MOID $<$ 1 LD among the Monte Carlo experiments. The first date of standard deviation in Mean Anomaly $>10^{\circ}$ is shown with initial orbit elements at the ephemeris retrieval date, 2022-09-15. $V_\infty$ is the relative velocity at the first time that MOID $<$ 1 LD of the nominal solution.}
\begin{tabular}{cccccccc}
\hline
NEO                & P(yr$^{-1}$) & $\Delta t/T$ & $t_{S>10^{\circ}}$  & a (au) & e      & i (deg) & $V_\infty$ (kms$^{-1}$) \\ \hline
  7482 (1994 PC1) & 1.51e-04 & 0.978 & 2541 & 1.349 & 0.330 & 33.47 & 19.68\\ 
  4179 Toutatis (1989 AC) & 5.19e-05 & 0.336 & 2516 & 2.545 & 0.625 & 0.45 & 12.19\\ 
314082 Dryope (2005 CZ36) & 4.88e-05 & 0.312 & 2371 & 2.238 & 0.575 & 16.14 & 14.05\\ 
 86819 (2000 GK137) & 4.44e-05 & 0.229 & 2563 & 1.996 & 0.506 & 10.06 & 10.07\\ 
143651 (2003 QO104) & 3.84e-05 & 0.306 & 2308 & 2.136 & 0.524 & 11.61 & 9.72\\ 
  5011 Ptah (6743 P-L) & 3.68e-05 & 0.152 & 2626 & 1.636 & 0.500 & 7.41 & 12.50\\ 
220839 (2004 VA) & 3.05e-05 & 0.172 & 2856 & 1.902 & 0.595 & 3.69 & 14.88\\ 
 66391 Moshup (1999 KW4) & 1.59e-05 & 0.045 & 2987 & 0.642 & 0.688 & 38.88 & 21.08\\ 
143404 (2003 BD44) & 1.45e-05 & 0.153 & 2587 & 1.968 & 0.606 & 2.66 & 15.93\\ 
190135 (2005 QE30) & 1.43e-05 & 0.074 & 2899 & 2.019 & 0.688 & 6.22 & 19.14\\ 
276732 (2004 EV9) & 1.07e-05 & 0.038 & 2809 & 1.471 & 0.781 & 40.83 & 32.09\\ 
 20236 (1998 BZ7) & 1.04e-05 & 0.087 & 2817 & 2.036 & 0.559 & 6.50 & 12.52\\ 
  4183 Cuno (1959 LM) & 7.37e-06 & 0.123 & 2913 & 1.982 & 0.636 & 6.67 & 17.01\\ 
387793 (2003 WL25) & 4.46e-06 & 0.030 & 2880 & 2.395 & 0.742 & 23.76 & 25.34\\ 
 29075 (1950 DA) & 4.22e-06 & 0.104 & 2913 & 1.698 & 0.508 & 12.17 & 14.09\\ 
214869 (2007 PA8) & 3.43e-06 & 0.091 & 2762 & 2.848 & 0.653 & 2.00 & 12.46\\ 
175114 (2004 QQ) & 2.30e-06 & 0.045 & 2648 & 2.249 & 0.664 & 5.72 & 19.74\\ 
       (2016 CB194) & 2.27e-06 & 0.039 & 2897 & 2.512 & 0.632 & 9.88 & 12.81\\ 
  7092 Cadmus (1992 LC) & 1.99e-06 & 0.046 & 2680 & 2.542 & 0.695 & 17.77 & 19.74\\ 
 90075 (2002 VU94) & 1.95e-06 & 0.082 & 2606 & 2.134 & 0.576 & 8.91 & 12.81\\ 
       (2019 HC) & 1.78e-06 & 0.015 & 2883 & 2.670 & 0.551 & 35.32 & 19.48\\ 
322966 (2002 KF4) & 1.64e-06 & 0.022 & 2960 & 2.903 & 0.577 & 37.02 & 19.43\\ 
  5143 Heracles (1991 VL) & 1.38e-06 & 0.032 & 2998 & 1.834 & 0.772 & 9.03 & 25.78\\ 
529718 (2010 KY127) & 1.32e-06 & 0.011 & 2908 & 2.489 & 0.883 & 60.84 & 39.67\\ 
508997 (2005 FL4) & 1.07e-06 & 0.012 & 2823 & 2.651 & 0.721 & 28.43 & 24.40\\ 
       (1999 XS35) & 5.39e-07 & 0.148 & 2409 & 17.780 & 0.948 & 19.62 & 18.28\\
       248590 (2006 CS) & 5.34e-07 & 0.005 & 2617 & 2.914 & 0.697 & 52.31 & 31.61\\ 
  1620 Geographos (1951 RA) & 4.47e-07 & 0.002 & 2861 & 1.246 & 0.335 & 13.34 & 11.88\\   \hline
\end{tabular}
\end{table}

Among the 40 km-sized NEOs currently with an Earth MOID $<$ 0.01 au, we find that their mean anomalies remain well defined typically for at least 200 years, and in some cases for thousands of years. On the other hand, there are a few examples of growth in mean anomaly uncertainty after 2200, such as 35396 (1997 XF11). Because the MOID becomes greater than 1 LD by the time the uncertainty in mean anomaly is large, the estimated probability for this NEO is zero.

The objects with the highest estimated probability are shown in Figure \ref{fig:upcoming-MA} and listed in Table \ref{tab:top10-hazLD}. The asteroid with the largest estimated probability of a deep close encounter is 7482 (1994 PC1). This result is to be expected, as in section \ref{sec:MOIDprop} and Figure \ref{fig:moid-frecs} we show that 7482 (1991 PC1) has a continuous low MOID. In this analysis we find that 7482 spends about 98\% of this millennium with an Earth MOID $<$1 LD. During this unusually lasting  MOID $<$1 LD period the position is well determined until approximately year 2500. 


The propagation of the MOID for the other km-sized NEOs on top of the list is shown in Figure \ref{fig:moid-hazardous-non0s}. We find that either the Earth MOID of these bodies is secularly drifting to zero, or that the current low-MOID oscillates around zero for a longer period of time. The latter case was observed for 7482 (1994 PC1), but additionally 4179 Toutatis (1989 AC) and 314082 Dryope (2005 CZ36) are in similar situations. Figure \ref{fig:moid-hazardous-non0s} shows that deep encounters are expected for these bodies, both in a low-MOID format and as the result of the Monte Carlo experiment.

The fact that the position is well determined allows us to determine the geometry of the subsequent close encounters, until the uncertainty grows too large. This assumption leaves a brief period of time between a very well constrained position and the range of validity of the uniformly distributed mean anomaly assumption. For this reason we check if there were actually such very close encounters among the low-MOID NEOs that we found earlier. In general, no close encounters within the Lunar Distance were found in the deterministic part of the trajectories. 

\newpage
There are a few exceptions that should be mentioned: 4179 Toutatis (1989 AC), 220839 (2004 VA) (Both in Figure \ref{fig:moid-hazardous-non0s}), 20236 (1998 BZ7), 214869 (2007 PA8) and 175114 (2004 QQ) experience close encounters right before or right after the date in which $S($MA$)>10^{\circ}$. In all of these cases, the MOID tends to zero around the dates in which a deep encounter is expected. In some cases, uncertainty in the position grows largely due to preceding close encounters. In general, we find that the Monte Carlo experiment agrees in finding deep encounters. Thus, the current method is successful in identifying their potential for very deep encounters within the next millennium.

\newpage
\section{Individual Hazard Analyses}\label{s:reports}

In this section we describe in more detail the hazardous nature of a few km-sized NEOs that were previously analyzed. We show the evolution of the MOID as well as the recorded track of close encounters. We study their orbital dynamics to provide context of the MOID evolution. In addition, we show the sequence of close encounters that precedes the growth in uncertainty and limits the accuracy of the prediction of the position of the NEO.

\subsection{Asteroid 7482 (1994 PC1)}

7482 (1994 PC1) has been highlighted in every section of this work because of its remarkable MOID evolution. Its Earth MOID is currently $6.09\cdot 10^{-4}$ au (0.237 LD), has been near zero for centuries and will remain very low for at least another 1000 years as shown in section \ref{sec:MOIDprop}. This condition is the reason why it is ranked as the most hazardous NEO in the list of Table \ref{tab:top10-hazLD}. 

The orbit elements of 7482 (1994 PC1) are shown for reference in Appendix \ref{App:Orbel}. During the period in which the Earth MOID remains small there are close encounters that cause significant variations semi-major axis and eccentricity. However, arguments of node and perihelion follow a secular drift. In its current orbit within the inner solar system, there is not a large amplitude of short-period oscillations that could disperse the distributions further. However, it is important to highlight that after 500 years the mean anomalies become uncertain. 

\begin{figure}[htb!]
	\centering
	\includegraphics[scale=1]{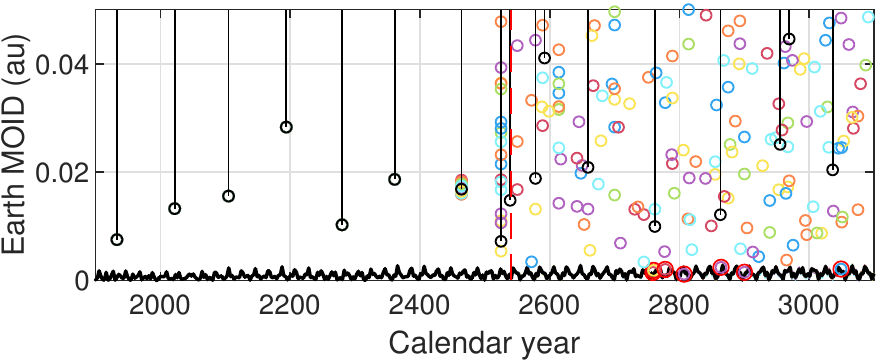}
	\hspace{3mm}
	\includegraphics[scale=1]{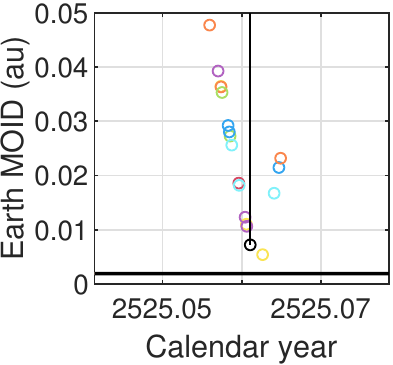}
	
	\caption{Earth close encounters of 7482 (1994 PC1), bounded by the osculating Earth MOID and as obtained through numerical Monte Carlo analysis. Close encounters are highlighted with vertical lines and points, and the Earth MOID is shown in continuous lines. Individual Monte-Carlo runs are shown in colors, black continuous line shows the nominal trajectory. Encounters of closest approach distance $< 1 LD$ (0.0026 au) are highlighted with a larger red circle.}
	\label{fig:prop-rep-PC1}
\end{figure}

Figure \ref{fig:prop-rep-PC1} shows the sequence of close encounters that are recorded in the Monte Carlo numerical propagation. It appears that the uncertainty in the encounter of 2525 is large enough that the range of possible closest approach distances is between 0 and 0.04 au. Right after the 2525 encounter the standard deviation in mean anomaly increases beyond 10 degrees, and we start estimating its probability of collision using the methods of section \ref{s:metric}. Encounters below the Lunar Distance were found after this period, which is consistent with the higher probability that we previously estimated.

\subsection{Asteroid  143651 (2003 QO104)}

The km-sized NEO with the shortest deterministic horizon is 143651 (2003 QO104), which was also previously introduced in Figure \ref{fig:upcoming-MA}. The orbit solution of 143651 (2003 QO104) has an observation arc of decades, including light-curve observations \citep{2009Birtwhistle} and radar astrometry \citep{2009Warner}. Thus, we believe that the rapid increase in uncertainty is a dynamical effect of its orbit. 

\begin{figure}[htb!]
	\centering
	\includegraphics[scale=1]{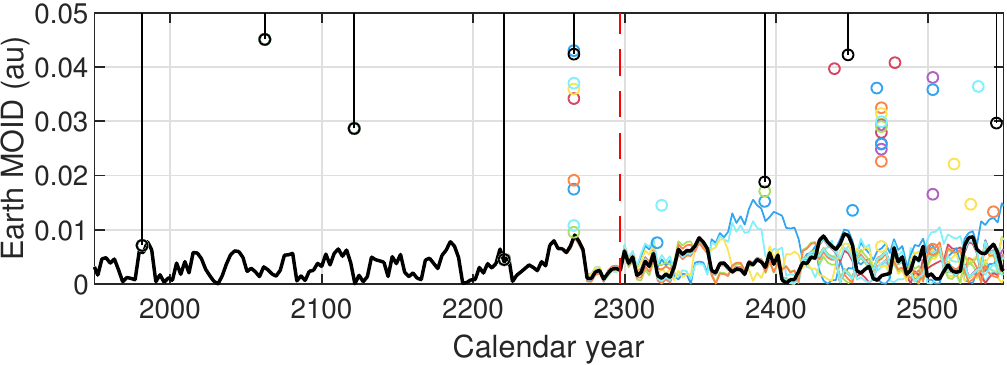}
	\caption{Earth close encounters of 143651 (2003 QO104), bounded by the osculating Earth MOID and as obtained through numerical Monte Carlo analysis. Close encounters are highlighted with vertical lines and points, and the Earth MOID is shown in continuous lines. Individual Monte-Carlo runs are shown in colors, black continuous line shows the nominal trajectory.}
	\label{fig:prop-rep-QO}
\end{figure}

Among the list of NEOs in Table \ref{tab:top10-hazLD} with non-zero estimated probability of having encounters below the Lunar Distance, 143651 (2003 QO104) has the slowest close encounters. These relative velocities imply larger scatter during close encounters, including a rapid increase in mean anomaly uncertainty. As shown in Figure \ref{fig:prop-rep-QO}, there is a close encounter in 2220 after which the sequence of encounters becomes unique for each Monte Carlo run.

\begin{figure}[htb!]
	\centering
	\includegraphics[scale=1]{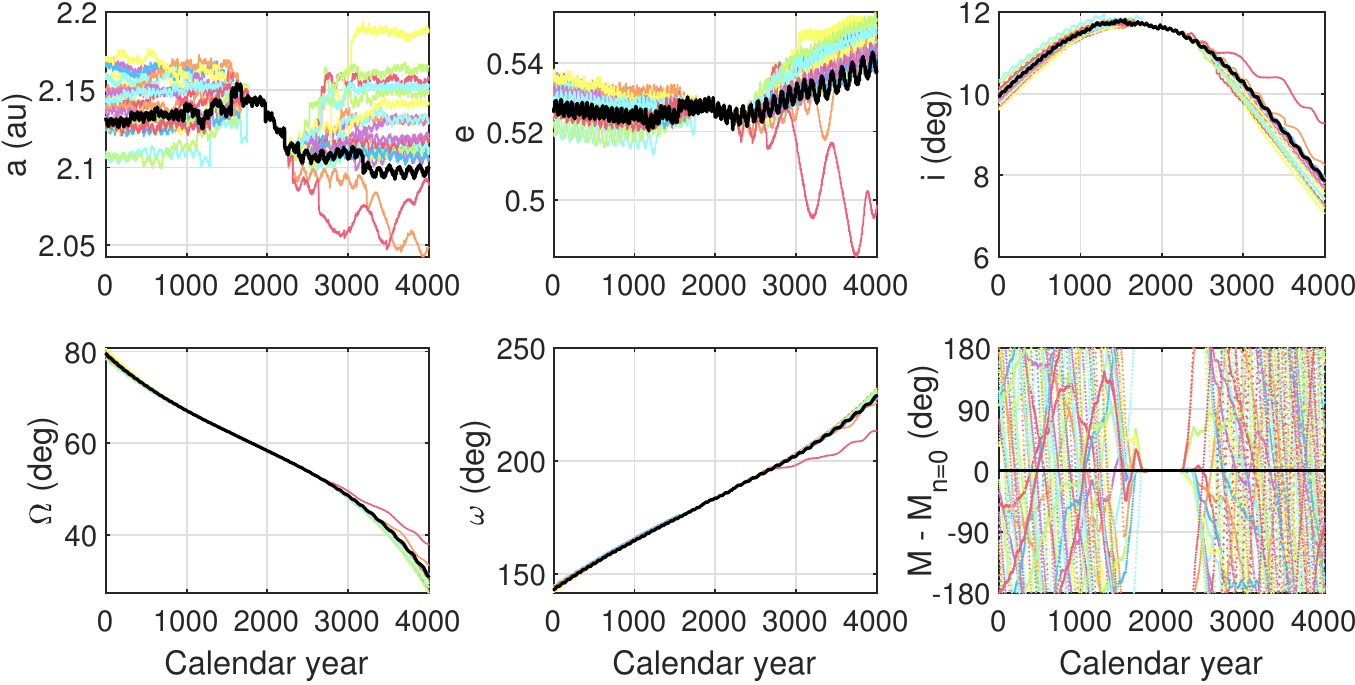}
	
	\caption{Numerical propagation of the orbit of 143651 (2003 QO104), a km-sized NEO. The trajectories are shown using the Keplerian elements semi-major axis, eccentricity, inclination, longitude of the ascending node, argument of perihelion and mean anomaly with respect to the nominal propagation. Individual Monte-Carlo runs (N=21) are shown in colors, the nominal trajectory is shown in a continuous black line.}
	\label{fig:prop-QO}
\end{figure}

Figure \ref{fig:prop-QO} shows the evolution of the orbital elements. By the end of the millennium there is a wide variety of orbits in which 143651 (2003 QO104), product of an undetermined sequence of both close and slow Earth close encounters.

\subsection{Asteroid 66391 Moshup (1999 KW4)}

The binary asteroid 66391 Moshup (1999 KW4) has been the object multiple studies relative to its binary system condition. It consists of a primary and secondary of respectively 1.317 km and 0.59 km of diameter \citep{Ostro_2006_science,SCHEIRICH2021114321}. Its rotation states suggest that it is a product of YORP spin-up and disruption \citep{Scheeres_2006_science,Davis_2020}, and its orbit is expanding in time due to the BYORP effect \citep{SCHEIRICH2021114321}.

\begin{figure}[htb!]
	\centering
	\includegraphics[scale=1]{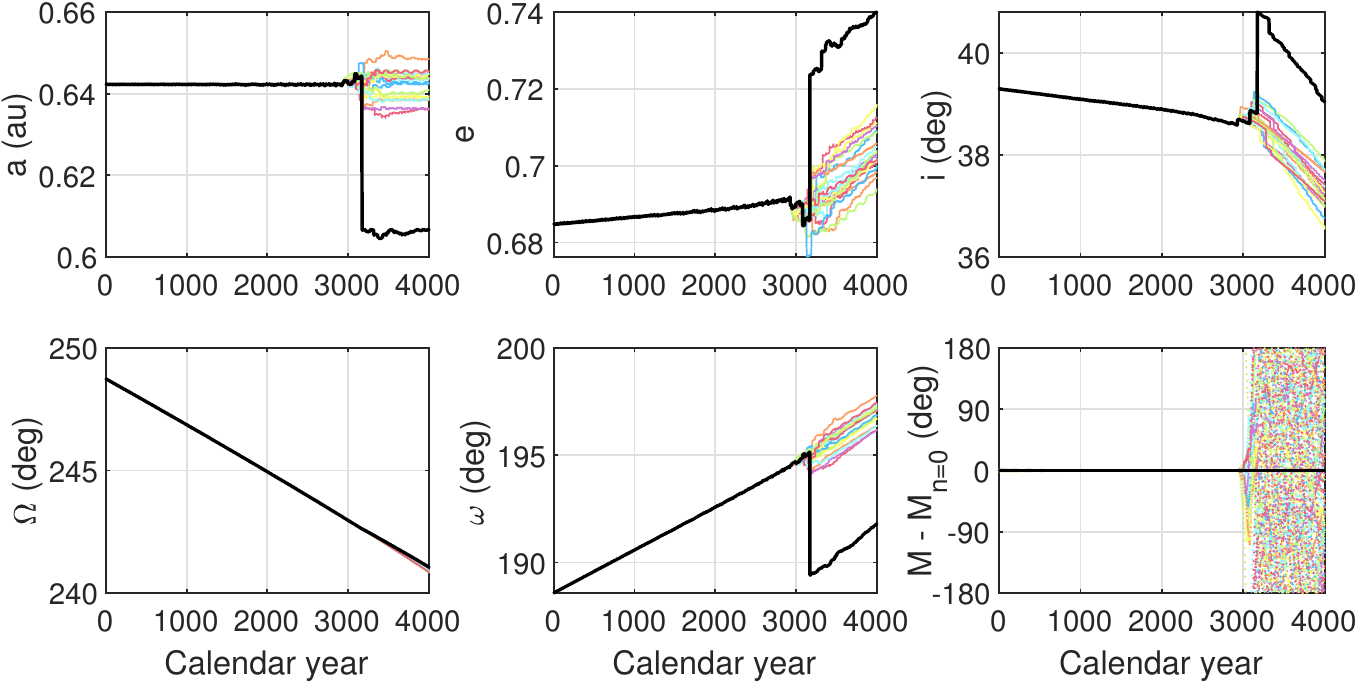}

	\caption{Numerical propagation of the orbit of 66391 Moshup (1999 KW4), a km-sized NEO. The trajectories are shown using the Keplerian elements semi-major axis, eccentricity, inclination, longitude of the ascending node, argument of perihelion and mean anomaly with respect to the nominal propagation. Individual Monte-Carlo runs (N=21) are shown in colors, the nominal trajectory is shown in a continuous black line.}
	\label{fig:prop-MO}
\end{figure}

The heliocentric orbit of 66391 Moshup (1999 KW4) is in resonance with the Earth, as it experiences resonant close encounters every 17 or 18 years. The  apparition of 2019 allowed observations from multiple observatories, during the 0.0346 au encounter \citep{SCHEIRICH2021114321}. The next close encounter will be in May 2036, with a closest distance of 0.0155 au, much closer than the first radar observations obtained using the Goldstone and Arecibo radar systems in May of 2001 \citep{Ostro_2006_science}. When the MOID becomes small, which is expected to happen slightly before year 3000, many close encounters below the Lunar Distance are recorded in our Monte Carlo analysis. These will cause a large scattering of the orbit as shown in Figure \ref{fig:prop-MO}. Because of how relatively late in the millennium the MOID $<$1LD condition is held, 66391 Moshup (1999 KW4) is not ranked higher in the list of Table \ref{tab:top10-hazLD}.

\begin{figure}[htb!]
	\centering
	\includegraphics[scale=1]{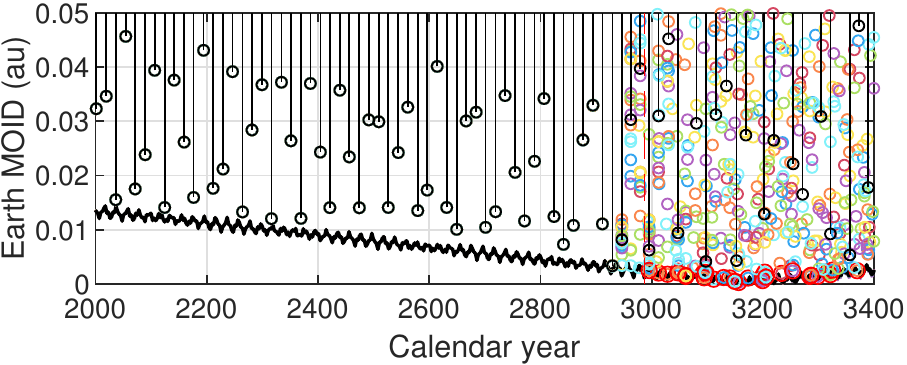}
	\hspace{3mm}
	\includegraphics[scale=1]{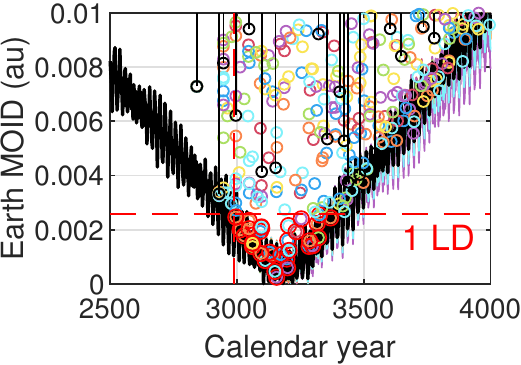}
	
	\caption{Earth close encounters of 66391 Moshup (1999 KW4), bounded by the osculating Earth MOID and as obtained through numerical Monte Carlo analysis. Close encounters are highlighted with vertical lines and points, and the Earth MOID is shown in continuous lines. Individual Monte-Carlo runs are shown in colors, black continuous line shows the nominal trajectory. Encounters of closest approach distance $< 1 LD$ (0.0026 au) are highlighted with a larger red circle.}
	\label{fig:prop-rep-Mosh}
\end{figure}

\subsection{Asteroid 29075 (1950 AD)}

Asteroid 29075 (1950 AD) is representative example of impact probability studies. After it was discovered and tracked for 17 days in 1950 \citep{wirtanen1950minor}, it was lost for 50 years until re-discovered on  2000-12-31. \cite{2002Giorgini} found a close approach in 2880 with the possibility of an impact. \cite{FARNOCCHIA2014321} modeled the Yarkovsky effect on 29075 (1950 AD) and estimated an impact probability of $2.5 \cdot 10^{-4}$.

\begin{figure}[htb!]
	\centering
	\includegraphics[scale=1]{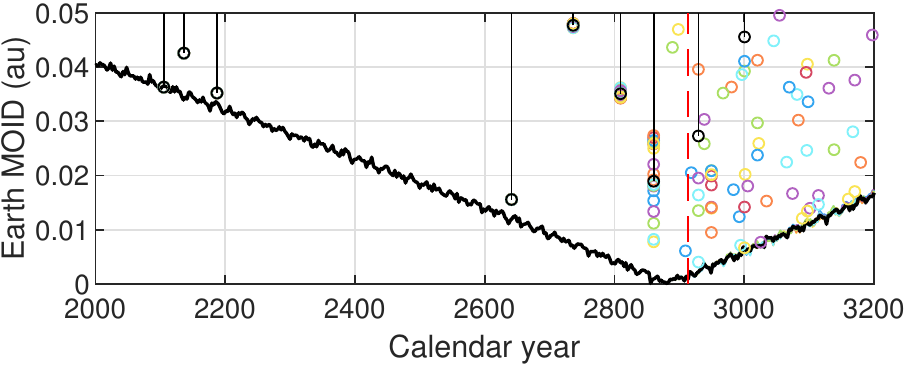}
	\hspace{3mm}
	\includegraphics[scale=1]{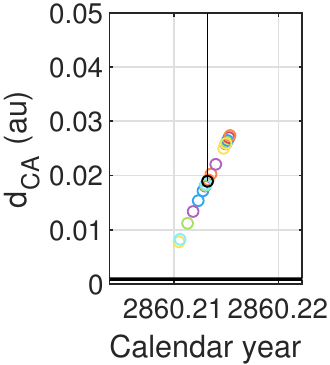}
	\includegraphics[scale=1]{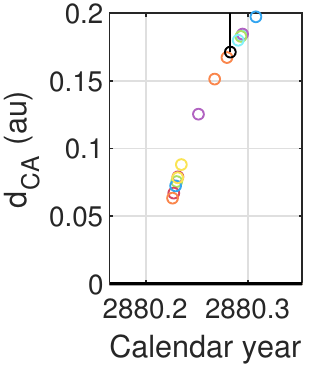}
	
	\caption{Earth close encounters of 29075 (1950 AD), bounded by the osculating Earth MOID and as obtained through numerical Monte Carlo analysis. Close encounters are highlighted with vertical lines and points, and the Earth MOID is shown in continuous lines. Individual Monte-Carlo runs are shown in colors, black continuous line shows the nominal trajectory.}
	\label{fig:prop-rep-AD}
\end{figure}

The example of 29075 (1950 AD) is paradigmatic in MOID evolution of NEOs. As shown in Figure \ref{fig:moid-examples}, its Earth MOID is secularly drifting to zero. During the decades that this condition is maintained, the probability of experiencing a deep encounter is non-zero. 29075 (1950 AD) was found among the list of km-sized NEOs for which we estimated this probability. As we show in Figure \ref{fig:prop-rep-AD}, the encounters of 2860 and 2880 will occur although with an uncertain closest approach distance.

\newpage
\subsection{Asteroid 2022 AP7}

The km-sized 2022 AP7 is one of the largest PHAs recently discovered \citep{Sheppard_2022}. The orbit of 2022 AP7 is in near-resonance with the orbit of the Earth, meaning that even if its MOID will become small in the next hundreds of years, almost no close encounters are expected in this period of time. The only likely exception is a close encounter in 2363 which will probably to be at a closest approach distance larger than 0.05au. An interesting finding is that 2022 AP7 comes from a sequence of resonant encounters every 5 years during the 19th century.

\begin{figure}[htb!]
	\centering

	\includegraphics[scale=1]{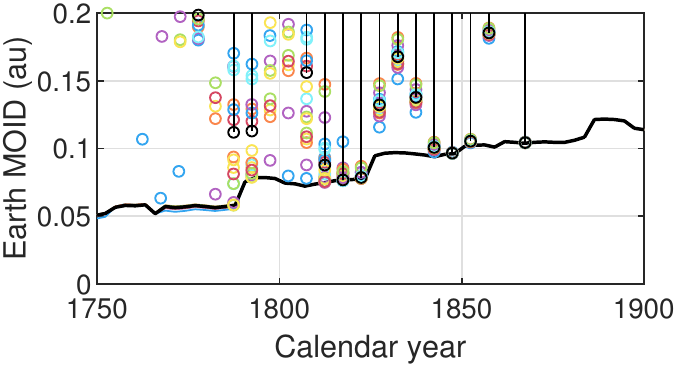}
        \hspace{3mm}
        \includegraphics[scale=1]{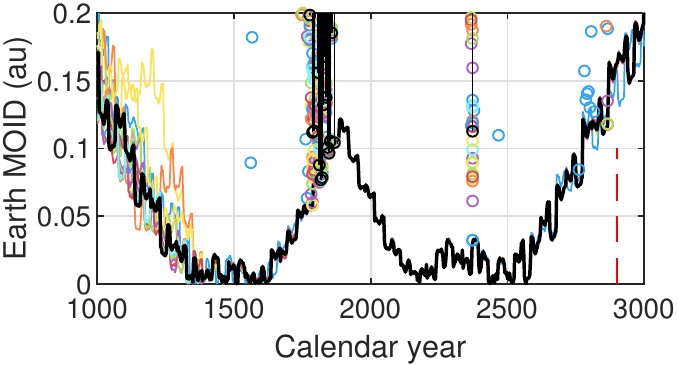}
        
	\caption{Earth close encounters of 2022 AP7, bounded by the osculating Earth MOID and as obtained through numerical Monte Carlo analysis. Close encounters are highlighted with vertical lines and points, and the Earth MOID is shown in continuous lines. Individual Monte-Carlo runs are shown in colors, black continuous line shows the nominal trajectory.}
	\label{fig:prop-rep-AP7}
\end{figure}


\section{Conclusions}


We characterized the long-term collision hazard of the known km-sized NEOs by the evolution of the MOID. The MOID can be accurately propagated beyond the dates in which the position within the orbit becomes unknown for certain NEOs. We first showed the km-sized NEOs with an Earth MOID $<0.01$ au of the next centuries. This classification already allowed us to rule out impacts for the majority of the population in the next 1000 years. When the position within the orbit is unknown and the MOID is small, we used an analytical estimation of the impact probability. We used this method to rank the km-sized population by the estimated probability of an Earth encounter of $d_{CA}<$ 1 LD. We found that there are a few km-sized near-Earth asteroids whose Earth MOID remains $<0.01$ au for thousands of years, such as 7482 (1994 PC1), 314082 Dryope (2005 CZ36), or 143651 (2003 QO104). 

In this work we push past the typical horizon for impact hazard assessment. Long-term impact hazard assessment can be limited by naturally chaotic dynamics. For example, the orbit of 143651 (2003 QO104) is scattered after a sequence of close encounters. However, in the cases in which the MOID can indeed be propagated confidently for thousands of years, we can point to the dates of interest for hazard characterization or rule out their risk.

As we propagate into larger orders of magnitude in time, it would be possible to simplify further our dynamics and use analytical \citep{Vokrouhlicky2012} or semi-analytical tools that accounted for the growth in uncertainty due close encounters \citep{Fuentes-Munoz2021}. The timescales of this study are long enough that the position is stochastic, but short enough that the precise modeling of the long-term effects is required. The range of orbits of the km-sized population allows widely different dynamical regimes. For these reasons, the use of numerical integration is left as the most reliable option. 

The metric derived in section \ref{s:metric} uses an analytical expression that assumes that the mean anomalies are uniformly distributed. This assumption holds when the uncertainty in mean anomaly is large, yet the transition between the deterministic part of the trajectory and this regime must be carefully analyzed. In some cases, these dates contribute the most to the probability of collision of the low-MOID period, as seen in the case of 29075 (1950 AD). We manually checked all the top-ranked asteroids for the presence encounters in this period of time, and displayed some of these examples in section \ref{s:reports}. The measure of the uncertainty in mean anomaly proves to be useful not only to validate the uniform distribution assumption, but to highlight dates of interest for hazard characterization. With this purpose in mind we find no need to increase the number of Monte Carlo samples to increase the accuracy in our predictions. The present work provides a list of asteroids and dates in which impact monitoring tools can be used to more accurately determine impact probabilities far beyond the default dates reported by impact monitoring systems.

Natural extensions of this work would be to broaden the selected group of asteroids from the km-sized population to PHAs or the whole NEO population. The MOID evolution as characterized in this work suggests a significant flux in and out of Earth’s vicinity, implying a significant flux in and out of the PHA category in timescales of decades to centuries. The long-term hazard ranking could be made available to the planetary defense community, as the most hazardous NEOs should be objects of interest for more detailed observations and future exploration missions.

\section{Acknowledgements}
This research is supported by grant 80NSSC22K0240 of the YORPD program of the National Aeronautics and Space Administration. Part of this research was carried out at the Jet Propulsion Laboratory, California Institute of Technology, under a contract with the National Aeronautics and Space Administration (80NM0018D0004).

\bibliography{references}
\bibliographystyle{aasjournal}

\newpage
\section{Appendix: Orbit elements propagation}\label{App:Orbel}

\begin{figure}[htb!]
	\centering
	\includegraphics[scale=1]{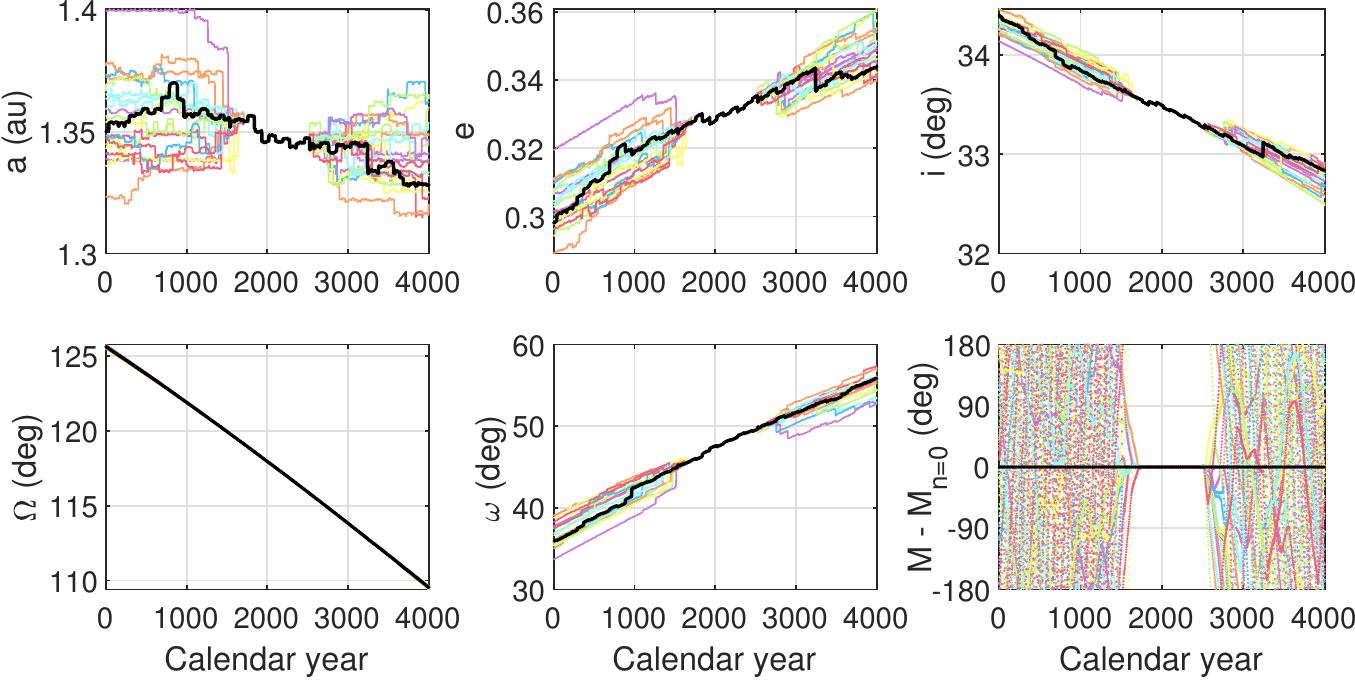}
	
	\caption{Numerical propagation of the orbit of 7482 (1994 PC1), a km-sized NEO. The trajectories are shown using the Keplerian elements semi-major axis, eccentricity, inclination, longitude of the ascending node, argument of perihelion and mean anomaly with respect to the nominal propagation. Individual Monte-Carlo runs (N=21) are shown in colors, the nominal trajectory is shown in a continuous black line.}
	\label{fig:prop-PC1}
\end{figure}

\begin{figure}[htb!]
	\centering
	\includegraphics[scale=1]{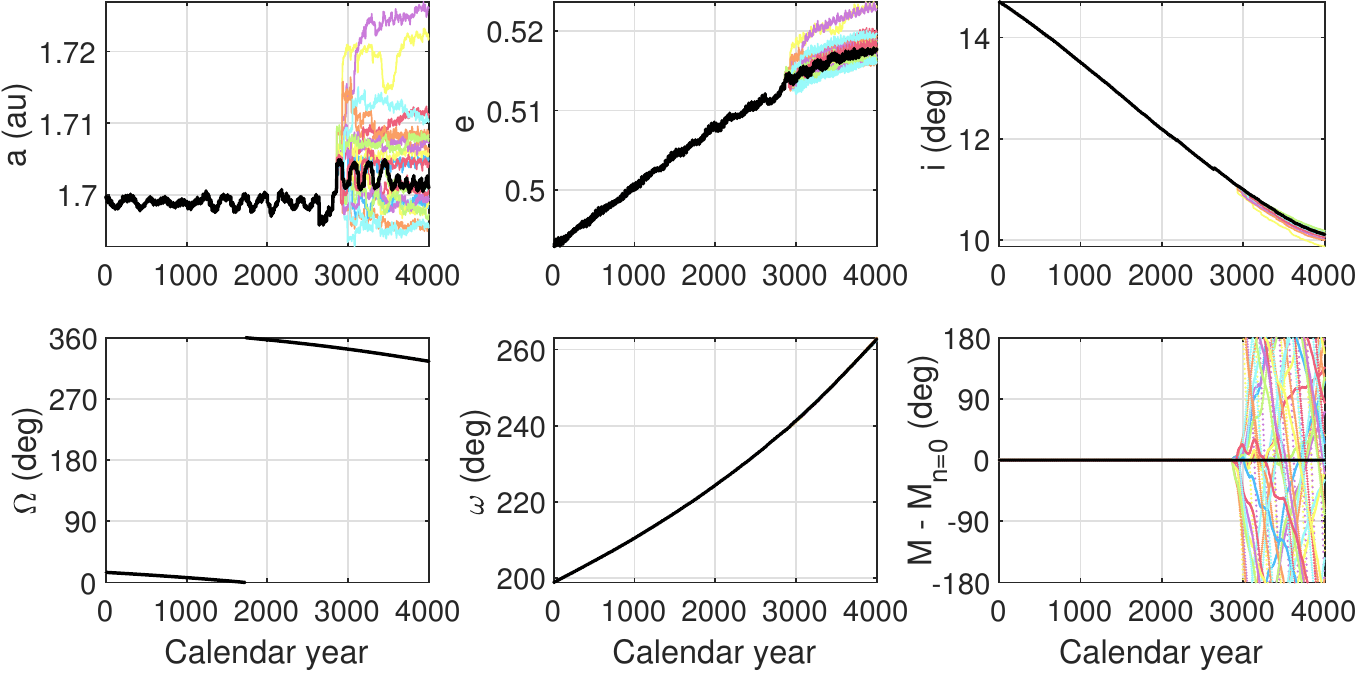}
	
	\caption{Numerical propagation of the orbit of 2022 AP7, a km-sized NEO. The trajectories are shown using the Keplerian elements semi-major axis, eccentricity, inclination, longitude of the ascending node, argument of perihelion and mean anomaly with respect to the nominal propagation. Individual Monte-Carlo runs (N=21) are shown in colors, the nominal trajectory is shown in a continuous black line.}
	\label{fig:prop-AD}
\end{figure}

\begin{figure}[htb!]
	\centering
	\includegraphics[scale=1]{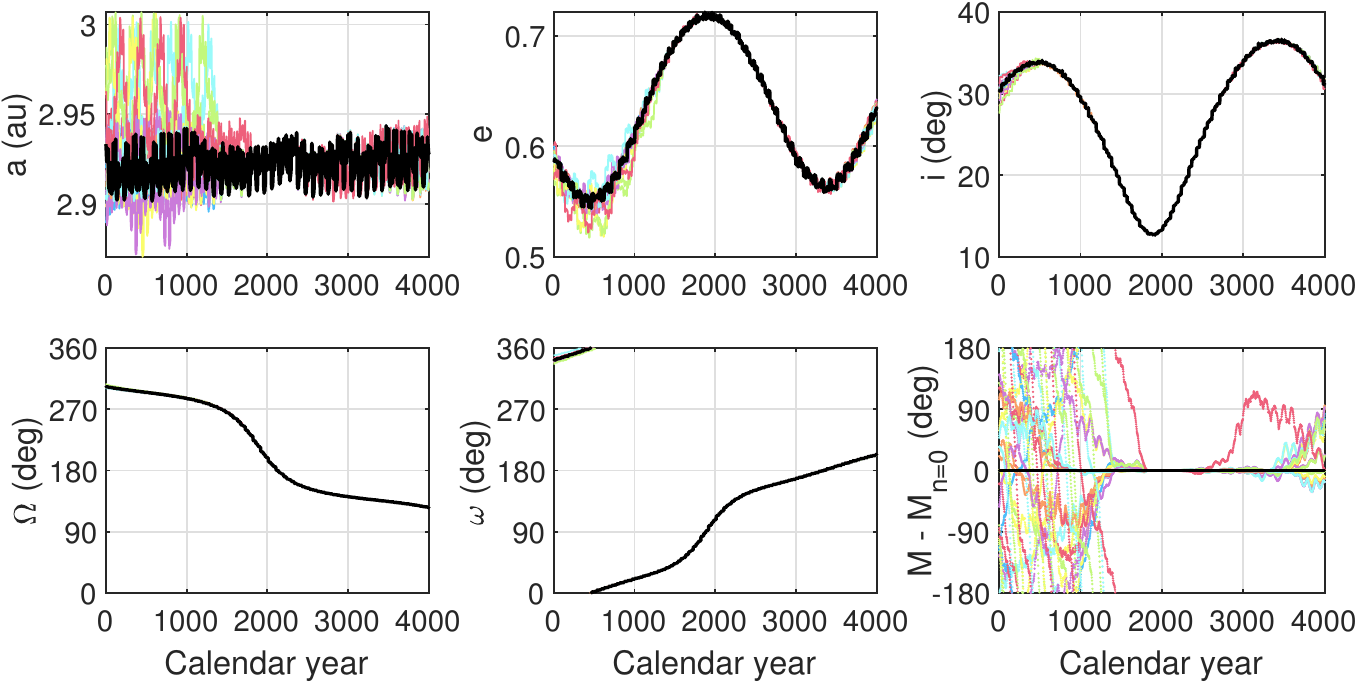}
	
	\caption{Numerical propagation of the orbit of 2022 AP7, a km-sized NEO. The trajectories are shown using the Keplerian elements semi-major axis, eccentricity, inclination, longitude of the ascending node, argument of perihelion and mean anomaly with respect to the nominal propagation. Individual Monte-Carlo runs (N=21) are shown in colors, the nominal trajectory is shown in a continuous black line.}
	\label{fig:prop-AP7}
\end{figure}





\end{document}